\documentclass[10pt,twocolumn]{article}

\usepackage[utf8]{inputenc}
\usepackage{amsmath}
\usepackage{amssymb}
\usepackage{titlesec}
\usepackage{cite}
\usepackage{booktabs}
\usepackage{graphicx}
\usepackage[colorlinks=false]{hyperref}
\usepackage[format=plain,labelfont=it]{caption}
\usepackage[left=1.5cm,right=1.5cm,top=2cm,bottom=2cm]{geometry}
\usepackage{xcolor}

\titleformat{\section}{\centering\normalfont\scshape}{\Roman{section}.}{5pt}{}
\titleformat{\subsection}{\normalfont\it}{\Alph{subsection}.}{5pt}{}
\titleformat{\subsubsection}{\normalfont\it}{\hspace{4mm}\arabic{subsubsection})}{5pt}{}

\newcommand\infoFootnote[1]{%
  \begingroup
  \renewcommand\thefootnote{}\footnote{#1}%
  \addtocounter{footnote}{-1}%
  \endgroup}

\newcommand{\preprintNoteIEEEAccepted}[1]{%
    © {#1} IEEE.  Personal use of this material is permitted.  Permission from IEEE must be obtained for all other uses, in any current or future media, including reprinting/republishing this material for advertising or promotional purposes, creating new collective works, for resale or redistribution to servers or lists, or reuse of any copyrighted component of this work in other works.}

\usepackage{pgfplots}
\pgfplotsset{compat=newest}
\usepackage{tikz}
\usetikzlibrary{arrows,positioning,patterns,decorations.pathreplacing}
\usepackage{tikz-3dplot}
\usetikzlibrary{arrows}
\usetikzlibrary{decorations.pathmorphing}
\usetikzlibrary{decorations.shapes}
\usetikzlibrary{decorations.text}
\usetikzlibrary{positioning, shapes}

\newtheorem{thm}{Theorem}

\newtheorem{lem}[thm]{Lemma}
\newtheorem{assum}{Assumption}
\newtheorem{proof}{Proof}

\newcommand{\R}{\mathbb{R}} 
\newcommand{\N}{\mathbb{N}}

\newcommand{\Cb}{\boldsymbol{C}}

\newcommand{\yb}{\boldsymbol{y}}

\newcommand{\xb}{\boldsymbol{x}}
\newcommand{\ub}{\boldsymbol{u}}

\newcommand{\Hb}{\boldsymbol{H}}
\newcommand{\Fb}{\boldsymbol{F}}
\newcommand{\Gb}{\boldsymbol{G}}

\newcommand{\Pb}{\boldsymbol{P}}
\newcommand{\Ab}{\boldsymbol{A}}
\newcommand{\Bb}{\boldsymbol{B}}
\newcommand{\Eb}{\boldsymbol{E}}
\newcommand{\Lb}{\boldsymbol{L}}
\newcommand{\Mb}{\boldsymbol{M}}
\newcommand{\Nb}{\boldsymbol{N}}

\newcommand{\Phib}{\boldsymbol{\Phi}}
\newcommand{\Db}{\boldsymbol{D}}
\newcommand{\Ib}{\boldsymbol{I}}

\newcommand{\Vb}{\boldsymbol{V}}
\newcommand{\Wb}{\boldsymbol{W}}

\newcommand{\zerob}{\boldsymbol{0}}

\newcommand{\Abc}{\boldsymbol{\mathcal{A}}}
\newcommand{\Bbc}{\boldsymbol{\mathcal{B}}}
\newcommand{\Cbc}{\boldsymbol{\mathcal{C}}}

\newcommand{\Kbc}{\boldsymbol{\mathcal{K}}}
\newcommand{\Fbc}{\boldsymbol{\mathcal{F}}}
\newcommand{\Wbc}{\boldsymbol{\mathcal{W}}}
\newcommand{\Lbc}{\boldsymbol{\mathcal{L}}}
\newcommand{\Mbc}{\boldsymbol{\mathcal{M}}}
\newcommand{\Nbc}{\boldsymbol{\mathcal{N}}}
\newcommand{\Pbc}{\boldsymbol{\mathcal{P}}}

\newcommand{\Vbc}{\boldsymbol{\mathcal{V}}}

\newcommand{\xbt}{\tilde{\boldsymbol{x}}}
\newcommand{\ybt}{\tilde{\boldsymbol{y}}}

\newcommand{\Vbt}{\tilde{\boldsymbol{V}}}

\DeclareMathOperator*{\rank}{rank}

\newtheorem{Theorem}{Theorem}

\newcommand{\blind}[1]{\textcolor{white}{#1}}

\DeclareMathAlphabet\mathbfcal{OMS}{cmsy}{b}{n}

\def\tvdots{\vbox{\baselineskip=2pt \lineskiplimit=0pt \kern6pt \Gbox{.}\Gbox{.}\Gbox{.}}} 
\title{\bf On the design of stabilizing FIR controllers}

\author{Janis Adamek$^\star$, Nils Schl\"uter$^\star$, and Moritz Schulze Darup}%
\begin{document}
\maketitle
\thispagestyle{empty}
\pagestyle{empty}
	
\begin{abstract}
Recently, it has been observed that finite impulse response controllers are an excellent basis for encrypted control, where privacy-preserving controller evaluations via special cryptosystems are the main focus. Beneficial properties of FIR filters are also well-known from digital signal processing, which makes them preferable over infinite impulse response filters in many applications. Their appeal extends to feedback control, offering design flexibility grounded solely on output measurements. However, designing FIR controllers is challenging, which motivates this work.  
To address the design challenge, we initially show that FIR controller designs for linear systems can equivalently be stated as static or dynamic output feedback problems. After focusing on the existence of stabilizing FIR controllers for a given plant, we tailor two common design approaches for output feedback to the case of FIR controllers. Unfortunately, it will turn out that the FIR characteristics add further restrictions to the LMI-based approaches. Hence, we finally turn to designs building on non-convex optimization, which provide satisfactory results for a selection of benchmark systems.
\end{abstract}
\infoFootnote{$^\star$Janis Adamek and Nils Schl\"uter share the first authorship.}
\infoFootnote{J. Adamek, N. Schl\"uter, and M. Schulze Darup are with the Department of Mechanical Engineering, TU Dortmund University, 44227 Dortmund, Germany. 
E-mails:  \{janis.adamek, nils.schlueter, moritz.schulzedarup\}@tu-dortmund.de.
} 
\infoFootnote{\hspace{-1.5mm}$^\ast$\preprintNoteIEEEAccepted{\textcolor{red}{2024}}
}
	
\section{Introduction}

Finite impulse response (FIR) filters are well-established and widely used in signal processing \cite{oppenheim2001discrete}.
In contrast, FIR feedback controllers (with orders larger than~$0$) are relatively rare. However, it has recently been observed that FIR controllers are quite useful in the framework of encrypted control \cite{Schlueter2021_ECC}. 
This young but emerging field of research deals with privacy-preserving networked control despite honest-but-curious platforms (e.g., clouds) or neighboring agents (see \cite{darup2021sruvey,marcolla2022survey} for an overview). More precisely, techniques such as homomorphic encryption 
are used to realize encrypted controller evaluations. While encrypted controllers offer many promising features, they are computationally demanding and subject to technical restrictions. For instance, an unlimited operating time is challenging to ensure for controllers with infinite impulse response (IIR). In contrast, it is straightforward and efficient to encrypt and operate FIR controllers using state-of-the-art homomorphic encryption schemes \cite{Schlueter2021_ECC}. Theoretical insights on this fact can be found in \cite{Schlueter2022_TAC}.

Regarding the design of FIR controllers (and filters), one can distinguish direct and indirect approaches. The latter typically build on IIR controllers (or filters) and consider FIR approximations, e.g., via frequency response sampling or windowing~\cite{diniz2010digital}.  However, FIR
approximations tend to have high orders, and they are not suited to describe unstable dynamics, which are well-known disadvantages. Nevertheless, an indirect approach has also been utilized to realize the encrypted FIR controllers in \cite{Schlueter2021_ECC}.

Direct FIR controller designs are rare and often restrictive.
 For instance, an $H_{\infty}$-based design using heuristics is proposed in \cite{Lee2005_FIR} 
and a convexified design for linear systems with uncertainties is considered in \cite{tregouet2011periodic}. 
In the context of encrypted control,
 \cite{tavazoei2023sufficient} derives a sufficient condition for the existence of FIR controllers for SISO systems, which can be verified using linear programming. 
 
Given the limitations of existing design approaches, especially for the MIMO systems considered here, this paper investigates whether FIR controllers may benefit from tailoring established designs for output feedback \cite{syrmos1997static}. 
The origin of this approach stems from our observation that FIR controller designs can be equivalently formulated as static or dynamic output feedback problems (as specified in Sect.~\ref{sec:specification}).
Hence, classical design approaches based on (convex) linear matrix inequalities (LMI)  \cite{Crusius1999_Sufficient_LMI,scherer1997multiobjective} can, in principle, be used and may offer problem-specific solutions. Unfortunately, our analysis (see Sect.~\ref{sec:nofreelunch}) shows that the FIR specifications lead to further restrictions, which we formalize as a helpful reference. 
Given the predominantly adverse theoretical outcomes, we conclude with a numerical case study, which confirms our results but also shows that heuristic non-convex designs can provide satisfactory designs albeit with higher computational efforts.

\textit{Roadmap.}
Section~\ref{sec:specification} deals with the problem specification and reformulations of the FIR controller design.
The existence of stabilizing FIR controllers is analyzed in Section~\ref{sec:maintheory}. Then, convexification strategies for FIR designs are investigated in Section~\ref{sec:nofreelunch}.  
Finally, we provide a numerical case study and conclusions in  Sections~\ref{sec:mainnumerics} and~\ref{sec:conclusion}, respectively.
\section{Problem specification and reformulations}
\label{sec:specification}

We consider linear discrete-time systems of the form
\begin{subequations}
\label{eq:originalSystem}
    \begin{align}
        \xb(k+1)&=\Ab \xb(k) +\Bb \ub(k)  \\
        \yb(k)&=\Cb \xb(k) 
    \end{align}
\end{subequations}
with the time step $k\in\N$, system state $\xb(k) \in \R^n$, control input $\ub(k) \in \R^m$, and output $\yb(k)\in \R^p$.
Throughout the paper, we make the following assumptions.
\begin{assum}
\label{assum:ABstabEtc}
    The pair $(\Ab,\Bb)$ is stabilizable and $(\Ab,\Cb)$ is detectable. Further, $\rank(\Bb)=m$ and $\rank(\Cb)=p$.
\end{assum}

Now, our focus is on the design of a stabilizing FIR feedback controller of the form 
\begin{equation}
    \label{eq:uFIR}
    \ub(k)=\sum_{i=0}^{\ell} \Fb_{i} \yb(k-i),
\end{equation} 
where $\Fb_i
$ denotes a controller gain and $\ell \in\N$ refers to the FIR's order.
For $\ell=0$, the controller \eqref{eq:uFIR} results in classical static output feedback $\Fb_0\yb(k)$. Remarkably, also the cases with $\ell>0$ can be interpreted in terms of static output feedback. To this end and inspired by \cite{Lee2005_FIR}, we introduce the augmented states and outputs
\begin{align*}
	\xbt(k):=\begin{pmatrix}\xb(k)\\
		\yb(k-1)\\
		\vdots\\
		\yb(k-\ell)
	\end{pmatrix} \quad \text{and} \quad 
	\ybt(k):=\begin{pmatrix}\yb(k)\\
		\yb(k-1)\\
		\vdots\\
		\yb(k-\ell)
	\end{pmatrix}
\end{align*}
which lead to the state space representation
\begin{subequations}
    \label{eq:augmentedSystem}
    \begin{align}
		\xbt(k+1)&=\Abc_\ell\xbt(k)+\Bbc_\ell\ub(k)\\
		\ybt(k)&=\Cbc_\ell\xbt(k)
	\end{align}
\end{subequations}
where the augmented matrices are given by
\begin{subequations}
\label{eq:augmentedSystemMatrices}
	\begin{align}
		\label{eq:augmentedSystemMatricesAB}
		\Abc_\ell&:=\begin{pmatrix}
			\Ab& \zerob&\zerob \\
			\Cb& \zerob&  \zerob \\
		\zerob& \Ib_{(\ell-1)p}&  \zerob \\
		\end{pmatrix}, & \Bbc_\ell &:=\begin{pmatrix}
			\Bb\\
			\zerob \\
			\zerob
		\end{pmatrix}, \\
		\quad\Cbc_\ell&:=\begin{pmatrix}
			\Cb& \zerob  \\
			\zerob & \Ib_{\ell p} 
		\end{pmatrix}.
	\end{align}
\end{subequations}
Clearly, for the augmented system~\eqref{eq:augmentedSystem}, the FIR controller~\eqref{eq:uFIR} results in static output feedback of the form
\begin{equation}
	\label{eq:uOutputFeedbackAugmentedY}
	\ub(k)=\Fbc_\ell \ybt(k) \;\; \text{with} \;\; \Fbc_\ell:=\begin{pmatrix}
	\Fb_0& \Fb_1 & \dots& \Fb_\ell
\end{pmatrix}.
\end{equation}
Now, instead of augmenting the system, one can also reformulate~\eqref{eq:uFIR} in terms of the linear dynamic controller 
\begin{subequations}
\label{eq:dynamicController}
    \begin{align}
	\hat{\xb}(k+1)&=\Hb \hat{\xb}(k) +\Gb \yb(k)  \\
	\ub(k)&=\Eb \hat{\xb} (k) + \Db \yb(k), 
	\end{align}
\end{subequations}
where $\hat{\xb}$ is the controller state and $\Hb$, $\Gb$, $\Eb$, and $\Db$ denote the controller parameters. In fact, specifying
\begin{subequations}
\label{eq:controllerParameters}
    \begin{align}
        \Hb&=\begin{pmatrix}
        \zerob&  \zerob \\
        \Ib_{\ell p}&  \zerob
        \end{pmatrix}, & 
        \Gb &=\begin{pmatrix}
        \Ib_p\\
        \zerob 
        \end{pmatrix}, \\
        \quad\Eb &=\begin{pmatrix}
        \Fb_1 & \dots& \Fb_\ell 
        \end{pmatrix}, \quad 
        & \Db&=\Fb_0.
    \end{align}
\end{subequations}
renders~\eqref{eq:dynamicController} equivalent to~\eqref{eq:uFIR}. 
Both reformulations allow specifying our focus on stabilizing FIR controllers with a given order $\ell$. In fact, based on~\eqref{eq:augmentedSystem}, we are aiming for $\Fbc_\ell$ leading to Schur stable  
\begin{equation}
\label{eq:closedLoopAugmented}
    \Abc_\ell+\Bbc_\ell\Fbc_\ell\Cbc_\ell.
\end{equation}
Analogously, building on~\eqref{eq:dynamicController}, we are looking for $\Eb$ and $\Db$ resulting in a Schur stable matrix
\begin{equation}
\label{eq:closedLoopDynamic}
\begin{pmatrix}
    \Ab+\Bb \Db \Cb & \Bb \Eb \\
    \Gb \Cb & \Hb
\end{pmatrix}.
\end{equation}
Remarkably, the closed-loop matrices \eqref{eq:closedLoopAugmented} and \eqref{eq:closedLoopDynamic} are equal. 
We formalize this observation in the following lemma
in order to simplify the discussion below.
\begin{lem}
\label{lem:equalClosedLoop}
	The matrices \eqref{eq:closedLoopAugmented} and \eqref{eq:closedLoopDynamic} specified by  \eqref{eq:augmentedSystemMatrices}, \eqref{eq:uOutputFeedbackAugmentedY}, and \eqref{eq:controllerParameters} are equal.
\end{lem}	
\begin{proof}
It is easy to see that both \eqref{eq:closedLoopAugmented} and \eqref{eq:closedLoopDynamic} result in
\begin{equation}
\label{eq:Phil}
\!\Phib_\ell:=\!\begin{pmatrix}
\Ab+\Bb \Fb_0 \Cb & \!\Bb \Fb_1\! & \dots& \!\Bb \Fb_{\ell-1}\! & \!\Bb \Fb_\ell \\
\Cb& \zerob&  \dots& \zerob&\zerob \\
\zerob& \Ib_p&  \dots& \zerob&\zerob \\
\vdots & \vdots & \ddots &\vdots & \vdots\\
\zerob & 	\zerob &  \dots & \Ib_p & \zerob
\end{pmatrix}\!,
\end{equation}
which immediately completes the proof.
\end{proof}

\section{Existence of stabilizing FIR controllers}
\label{sec:maintheory}

Before we dive into possible design strategies for~\eqref{eq:uFIR}, we derive a necessary condition for the existence of a stabilizing FIR controller. This condition builds on the trivial observation that all eigenvalues of $\Hb$ in \eqref{eq:controllerParameters} are zero. As a consequence, $\Hb$ is Schur stable and the dynamic controller in~\eqref{eq:dynamicController} is (asymptotically) stable. Since this controller is an equivalent representation of~\eqref{eq:uFIR}, we conclude that the FIR controller design can only be successful if the system is stabilizable by a stable controller. Such systems are called strongly stabilizable and they possess the parity interlacing property (PIP) \cite{vidyasagar2011control}.
Now, for single-input single-output (SISO) systems, we find the following condition.

\begin{Theorem}(\cite[(PIP-SS)]{hagiwara1988preservation})
   \label{th:strongstab}
   A discrete-time SISO system has the PIP 
   if and only if 
   the number of positive real unstable poles between two positive real unstable zeros is even.
\end{Theorem}
Variants of this theorem also exist for multiple-input multiple-output (MIMO) systems but, to the best of the authors' knowledge, only for the continuous-time case \cite[p. 84]{vidyasagar2011control}. While a technical extension to the discrete-time MIMO case is beyond the scope of this paper, it is still clear by definition that strong stabilizability is a necessary condition for the existence of a stabilizing FIR controller.
In order to highlight this point and for later validation, we consider the two SISO systems
\begin{align}
\label{eq:PIPexamples}
   G_1(z)=\frac{z-2}{(z-3)(z-4)} \,\,\,\, \text{and} \,\,\,\,
   G_2(z)=\frac{z-2}{z\left(z-3\right)},
\end{align}
which we specify via transfer functions for convenience. Taking into account that $\infty$ counts as an unstable zero for strictly proper systems, it is easy to see that $G_1$ possesses the PIP whereas $G_2$ does not. 
Now, while only $G_1$ is strongly stabilizable, classical stabilizability is given for both systems. For instance, the controllers 
\begin{equation*}
    C_1(z)=\frac{5.6 z + 0.1}{z} \quad \text{and} \quad   C_2(z)=\frac{39z}{z^2+4-26} 
\end{equation*}
stabilize $G_1$ and $G_2$, respectively. Here, $C_1$ is a (stable) FIR controller, whereas $C_2$ is neither FIR nor stable. In fact, due to Theorem~\ref{th:strongstab}, it is impossible to find a stabilizing (and by construction stable) FIR controller for $G_2$ (as later confirmed by our numerical studies in Section~\ref{sec:mainnumerics}).

Next, we motivate that strong stabilizability is not only necessary but also sufficient for the existence of stabilizing FIR controllers. 
In fact, strong stabilizability guarantees the existence of a stable dynamic controller of the form~\eqref{eq:dynamicController} but without the structural restrictions in~\eqref{eq:controllerParameters}. 
Nonetheless, it is well-known that such dynamics with a Schur stable $\Hb$ can be approximated with arbitrary precision by FIR dynamics. 
To see this, note that the output of~\eqref{eq:dynamicController} for an initial controller state  $\hat{\xb}(0)=\zerob$ is given by 
\begin{equation}
    \label{eq:dynamicOutput}
    \sum_{i=0}^{k-1} \Eb \Hb^i \Gb\yb(k-1-i) + \Db \yb(k).
\end{equation}
Computing FIR approximations using a rectangular window method leads to the specifications  $\Fb_0 :=\Db$ 
and $\Fb_i := \Eb \Hb^{i-1} \Gb$ for every $i \in \{1,\dots,\ell\}$ (see~\cite[Chapter 5.4]{diniz2010digital}).
Then, the difference between~\eqref{eq:dynamicOutput} and~\eqref{eq:uFIR} is 
$\sum_{i=\ell}^{k-1} \Eb \Hb^i \Gb\yb(k-1-i)$.
Clearly, for sufficiently large FIR orders $\ell$, the difference tends to $0$ (and the same observation holds for $\hat{\xb}(0) \neq \zerob$ if also $k$ is large enough). As a consequence, stable dynamic controllers suggest the existence of stabilizing FIR controllers. Nevertheless, in this paper, we aim for a direct design of FIR controllers and not for the approximation of stable IIR controllers, since this often requires unnecessarily large orders $\ell$. 
Still, large $\ell$ can be beneficial for a direct design, as indicated by the following lemma. In fact, the feasibility of a controller design may only be obtained but never lost when increasing~$\ell$.

\begin{lem}
\label{lem:l+1}
If there exists a $\Fbc_\ell$ such that the closed-loop matrix $\Phib_\ell$ is Schur stable for some $l\in \N$, then there also exists a $\Fbc_{\ell+1}$ such that $\Phib_{\ell+1}$ is Schur stable.
\end{lem}
\begin{proof}
 The special choice $\Fbc_{\ell+1}=\begin{pmatrix}
     \Fbc_\ell& \zerob
 \end{pmatrix}$ leads to the closed-loop matrix 
 $$
 \Phib_{\ell+1}=\begin{pmatrix}
     \Phib_\ell  & \zerob\\
     \begin{pmatrix}\zerob & \Ib_p\end{pmatrix} & \zerob 
 \end{pmatrix} \in \R^{(n+(\ell+1)p) \times (n+(\ell+1)p)}.
 $$ 
 As apparent from the block-triangular structure with a zero-block on the diagonal,
 $\Phib_{\ell+1}$ inherits all eigenvalues of the Schur stable matrix $\Phib_\ell$ and additionally has $p$ stable eigenvalues at $0$, which immediately proofs the claim. 

\end{proof}

In summary, designing stabilizing FIR controllers is only meaningful for strongly stabilizable systems and for these, it is reasonable to search for the smallest feasible FIR order $\ell$.

\section{No free lunch from classical approaches}
\label{sec:nofreelunch}

A stabilizing FIR controller is characterized by a Schur stable matrix $\Phib_\ell$, as in~\eqref{eq:Phil}. Clearly, based on Lyapunov's theory, Schur stability holds if and only if there exists a positive definite $\Pbc_\ell$ satisfying 
\begin{equation}
    \label{eq:dlyapFC}
    \Phib_\ell^{\top} \Pbc_\ell \Phib_\ell- \Pbc_\ell \prec \zerob.
\end{equation}
Unfortunately, finding suitable $\Fbc_\ell$ and $\Pbc_\ell$ or even deciding whether they exist or not is significantly harder than solving the related problem 
\begin{equation}
\label{eq:dlyapK}
(\Abc_\ell+\Bbc_\ell\Kbc_\ell)^{\top} \Pbc_\ell (\Abc_\ell+\Bbc_\ell\Kbc_\ell)- \Pbc_\ell \prec \zerob
\end{equation}
associated with static state feedback (i.e., $\ub(k)=\Kbc_\ell \xbt(k)$).
In fact, the nonlinear coupling of $\Kbc_\ell$ and $\Pbc_\ell$ in~\eqref{eq:dlyapK} can be resolved by carrying out a congruence transformation with $\Wbc_\ell:=\Pbc_\ell^{-1}$, applying the Schur complement, and using the substitution $\Lbc_\ell=\Kbc_\ell \Pbc_\ell^{-1}$. Then, we obtain the~LMI
\begin{equation}
\label{eq:WLMIforK}
\begin{pmatrix}
\Wbc_\ell  &   \Abc_\ell\Wbc_\ell +\Bbc_\ell\Lbc_\ell \\
  \Wbc_\ell \Abc_\ell^\top +\Lbc_\ell^\top \Bbc_\ell^\top  & \Wbc_\ell
\end{pmatrix} \succ \zerob,
\end{equation}
which solves~\eqref{eq:dlyapK} by means of $\Kbc_\ell=\Lbc_\ell\Wbc_\ell^{-1}$.
Clearly, if the resulting $\Kbc_\ell$ is such that an $\Fbc_\ell$ exists which satisfies
\begin{equation}
\label{eq:FCK} 
\Fbc_\ell\Cbc_\ell=\Kbc_\ell,
\end{equation}
then $\Pbc_\ell$ and  $\Fbc_\ell$ also solve~\eqref{eq:dlyapFC}. Remarkably, a decomposable solution for~\eqref{eq:dlyapK} as in~\eqref{eq:FCK} exists whenever~\eqref{eq:dlyapFC} is solvable. This trivially follows from the fact that a solution to~\eqref{eq:dlyapFC} in terms of  $\Pbc_\ell$ and $\Fbc_\ell$  implies a solution to \eqref{eq:dlyapK} in terms of $\Pbc_\ell$ and $\Kbc_\ell=\Fbc_\ell\Cbc_\ell$, where $\Kbc_\ell$ is decomposable by construction.

Now, it is interesting to note that \eqref{eq:dlyapK} and consequently~\eqref{eq:WLMIforK} are always feasible for our setup. In fact, the feasibility of \eqref{eq:dlyapK} holds once the pair $(\Abc_\ell,\Bbc_\ell)$ is stabilizable, which is given according to the following lemma in combination with Assumption~\ref{assum:ABstabEtc}.
\begin{lem}
The pair $(\Abc_\ell,\Bbc_\ell)$ is stabilizable if and only if $(\Ab,\Bb)$ is stabilizable.
\end{lem}
\!\!\!\!\!\begin{proof}\!
By definition, $(\Ab,\Bb)$ is stabilizable if and only if
$$
\rank \begin{pmatrix}
    \lambda \Ib_n - \Ab & \Bb 
\end{pmatrix} = n
$$
for all unstable eigenvalues $\lambda$ of $\Ab$. 
Likewise,  $(\Abc_\ell,\Bbc_\ell)$ is stabilizable if and only if
$$
\rank \begin{pmatrix}
    \tilde{\lambda} \Ib_{n+\ell p} - \Abc_\ell & \Bbc_\ell
\end{pmatrix} = n+\ell p
$$
for all unstable eigenvalues $\tilde{\lambda}$ of $\Abc_\ell$. As apparent from \eqref{eq:augmentedSystemMatricesAB}, $\Abc_\ell$ inherits the $n$ eigenvalues of $\Ab$ and additionally has $\ell p$ stable eigenvalues at $0$. Hence, the unstable eigenvalues (if any) of $\Ab$ and $\Abc_\ell$ are equivalent. Now, by inspecting $(\tilde{\lambda} \Ib_{n+\ell p} - \Abc_\ell \quad \Bbc_\ell)$, it becomes clear that its rank is determined by the sum of the ranks of the two matrices
\begin{align}
\label{eq:Matrices_Hautus}
   \begin{pmatrix}
    \tilde{\lambda} \Ib_n - \Ab & \Bb 
\end{pmatrix} \quad \text{and} \quad  \begin{pmatrix}
				\tilde{\lambda} \Ib_p&  \dots& \zerob&\zerob  \\
				\Ib_p& \ddots   & \zerob&\zerob  \\
				\vdots & \ddots &\ddots & \vdots \\
				\zerob &  \dots & \Ib_p & \tilde{\lambda} \Ib_p 
			\end{pmatrix}. 
\end{align}
The second matrix in \eqref{eq:Matrices_Hautus} obviously offers a (full) rank of $\ell p$ for any unstable $\tilde{\lambda}$. The first matrix has the rank $n$ required for stabilizability if and only if $(\Ab,\Bb)$ is stabilizable.
\end{proof}
Given the guaranteed feasibility and convexity of \eqref{eq:WLMIforK}, one could aim for solving~\eqref{eq:WLMIforK} and trying to derive a solution to~\eqref{eq:dlyapFC} via~\eqref{eq:FCK}. 
This is expedient in special cases such as,
e.g., $\Cb=\Ib_n$. 
However, in general, the ability to find a suitable $\Fbc_\ell$ requires explicitly considering the constraint~\eqref{eq:FCK} while solving~\eqref{eq:dlyapK} or, equivalently, to solve~\eqref{eq:dlyapFC} directly. 
Unfortunately, both problems are typically hard~\cite{Toker1995_NPhardnessBMI}.
Hence, classical solution strategies 
build on different convexifications of~\eqref{eq:dlyapFC} that enable an efficient numerical solution. 
Two popular ones are discussed next.
\subsection{Convexification for static output feedback}
\label{subsec:staicconv}

A classical approach to solve~\eqref{eq:dlyapFC} with $\Phib_\ell$ as in \eqref{eq:closedLoopAugmented} first transforms~\eqref{eq:dlyapFC} into 
$$
\begin{pmatrix}
\Wbc_\ell  &   \Abc_\ell\Wbc_\ell +\Bbc_\ell \Fbc_\ell \Cbc_\ell \Wbc_\ell \\
  \Wbc_\ell \Abc_\ell^\top +\Wbc_\ell \Cbc_\ell^\top \Fbc_\ell^\top \Bbc_\ell^\top  & \Wbc_\ell
\end{pmatrix} \succ \zerob
$$
analogously to the transformations applied to~\eqref{eq:dlyapK} which resulted in~\eqref{eq:WLMIforK}. Then, as proposed in~\cite{Crusius1999_Sufficient_LMI} and \cite{tregouet2011periodic}, one replaces $\Cbc_\ell \Wbc_\ell$ with $\Mbc_\ell \Cbc_\ell$ and subsequently $\Fbc_\ell \Mbc_\ell$ with $\Nbc_\ell$, which yields the LMI
\begin{subequations}
\label{eq:convexAugmentedOutput}
\begin{align}
\!\begin{pmatrix}
\Wbc_\ell  &   \!\Abc_\ell\Wbc_\ell +\Bbc_\ell \Nbc_\ell \Cbc_\ell \\
\Wbc_\ell \Abc_\ell^\top +\Cbc_\ell^\top \Nbc_\ell^\top \Bbc_\ell^\top  & \Wbc_\ell
\end{pmatrix} \succ \zerob,&\\
\label{eq:convexAugmentedOutput_constraint}
\Mbc_\ell \Cbc_\ell =\Cbc_\ell \Wbc_\ell,&
\end{align}
\end{subequations}
where the first substitution involving $\Mbc_\ell \in \R^{(\ell+1)p \times (\ell+1)p}$ acts as a constraint. Now, if \eqref{eq:convexAugmentedOutput} is feasible, then $\Pbc_\ell=\Wbc_\ell^{-1}$ and $\Fbc_\ell=\Nbc_\ell \Mbc_\ell^{-1} $ solve~\eqref{eq:dlyapFC}. 
Unfortunately, the constraint \eqref{eq:convexAugmentedOutput_constraint} and the artificial structure $\Bbc_\ell \Nbc_\ell \Cbc_\ell$ are restrictive, which often results in infeasibility of~\eqref{eq:convexAugmentedOutput} even if~\eqref{eq:dlyapFC} has a solution. 
This observation can be formalized for FIR controller design. 
In fact, according to the following lemma, increasing the FIR order is not helpful for this approach if a design via \eqref{eq:convexAugmentedOutput} fails for $\ell=0$, i.e., for classical static output feedback.

\begin{lem}
    \label{lem:ConvexInfeasible}
    If \eqref{eq:convexAugmentedOutput} is infeasible for $\ell=0$, it is infeasible for any $\ell \in \N$.
\end{lem}

\begin{proof}
In order to prove the claim, we partition the variables in \eqref{eq:convexAugmentedOutput} according to 
$$
 \Wbc_{\ell}:=\!\begin{pmatrix}
            \Wb_{0} &  \ast \\
            \ast &  \ast
        \end{pmatrix}\!, \,\, \Mbc_{\ell}:=\!\begin{pmatrix}
            \Mb_{0} &  \ast \\
            \ast &  \ast
        \end{pmatrix}\!,\,\,
        \Nbc_\ell:=\!\begin{pmatrix}
            \Nb_{0}  & \ast 
        \end{pmatrix},
$$
where only the blocks $\Wb_0 \in \R^{n\times n}$, $\Mb_0 \in \R^{p\times p}$, and $\Nb_0 \in \R^{n\times p}$ are relevant for the proof. 
 Feasibility of \eqref{eq:convexAugmentedOutput} for $\ell=0$  then requires 
 \begin{subequations}
 \label{eq:convexAugmentedOutput_0}
  \begin{align}
  \label{eq:convexAugmentedOutput_W}
        \begin{pmatrix}
            \Wb_{0}&\Ab\Wb_{0}+\Bb\Nb_{0}\Cb\\
            \Wb_{0} \Ab^\top \!\!+\Cb^\top \Nb_0^\top \Bb^\top  & \Wb_{0}
        \end{pmatrix} \succ \zerob,& \\
        \label{eq:convexAugmentedOutput_M}
    \Mb_0 \Cb =\Cb \Wb_0.&
        \end{align}
 \end{subequations}
Based on the partitions above and the augmented system matrices in~\eqref{eq:augmentedSystemMatrices}, the conditions for $\ell>0$ 
read
  \begin{align*}
  \begin{pmatrix}
            \Wb_{0} &  \ast &\Ab\Wb_{0}+\Bb\Nb_{0}\Cb & \ast\\
            \ast &  \ast & \ast & \ast \\
            \Wb_{0} \Ab^\top \!\!+\Cb^\top \Nb_0^\top \Bb^\top  & \ast &  \Wb_{0} & \ast \\
            \ast & \ast & \ast & \ast 
        \end{pmatrix} \succ \zerob,& \\
    \begin{pmatrix}
       \Mb_0 \Cb & \ast \\
       \ast & \ast
    \end{pmatrix} =\begin{pmatrix}
       \Cb \Wb_0 & \ast \\
       \ast & \ast
    \end{pmatrix}.&
        \end{align*}
 The equality conditions obviously contain~\eqref{eq:convexAugmentedOutput_M}. 
By noting that all 
 principal submatrices of a positive definite matrix are likewise positive definite (see, e.g., \cite[Prop.~1.56]{Bai_Matrix_Analysis}), the LMI also contains \eqref{eq:convexAugmentedOutput_W}. Hence, the conditions for $\ell>0$ cannot be fulfilled if~\eqref{eq:convexAugmentedOutput_0} is infeasible.
 \end{proof}

\subsection{Convexification for dynamic output feedback}
\label{subsec:dynamicconvex}

Another widely applied approach, which can, e.g., be found in~\cite{scherer1997multiobjective}, is tailored towards the design of a dynamic output feedback controllers in the form~\eqref{eq:dynamicController} but without the structural restrictions in~\eqref{eq:controllerParameters}. 
In other words, all controller parameters are variable here apart from a predefined controller dimension.  
The approach likewise considers a congruence transformation of~\eqref{eq:dlyapFC} with $\Wbc_\ell$ and a subsequent application of the Schur complement. 
However, it exploits the structure \eqref{eq:closedLoopDynamic} of $\Phib_\ell$ (instead of \eqref{eq:closedLoopAugmented} as before). Moreover, it involves another congruence-like transformation with a full column rank matrix $\Vbc_\ell$ (specified below), which leads to
\begin{align}
\label{eq:W-Problem_dyn}
\begin{pmatrix}
\Vbc_\ell^\top & \zerob \\ \zerob & \Vbc_\ell^\top
\end{pmatrix}
\begin{pmatrix}
\Wbc_\ell  &   \Phib_\ell \Wbc_\ell  \\
\Wbc_\ell \Phib_\ell^{\top}  & \Wbc_\ell
\end{pmatrix} 
\begin{pmatrix}
\Vbc_\ell & \zerob \\ \zerob & \Vbc_\ell
\end{pmatrix}
\succ \zerob.
\end{align}
Now, the idea is to carefully select $\Vbc$ such that both a reversible replacement of the controller variables and a convexification of~\eqref{eq:W-Problem_dyn} is achieved. 
To this end, one introduces the partitions 
\begin{align*}
    \Wbc_\ell:=\begin{pmatrix}
    \Wb_{0} & \Wb_1\\ \Wb_1^\top & \Wb_2
    \end{pmatrix}   \quad \text{and} \quad
    \Wbc_\ell^{-1}:=\begin{pmatrix}
    \Pb_{0} & \Pb_1\\ \Pb_1^\top & \Pb_2
    \end{pmatrix}, 
\end{align*}
where the block matrices are conformal with~\eqref{eq:closedLoopDynamic}, and specifies the matrix
\begin{equation}
    \label{eq:VScherer}
     \Vbc_\ell:=\begin{pmatrix}
        \Pb_0 & \Ib_n \\ \Pb_1^\top &\zerob
    \end{pmatrix}.
\end{equation}
Due to $\Wb_{0} \Pb_0+\Wb_1 \Pb_1^\top=\Ib_n$, $\Wb_{1}^\top \Pb_0+\Wb_2 \Pb_1^\top=\zerob$, and the symmetry of $\Wb_0$ and $\Pb_0$, we find
$$
    \Vbc_\ell^\top \Wbc_\ell \Vbc_\ell=
    \begin{pmatrix}
        \Pb_0 & \Pb_1 \\ \Ib_n & \zerob        
    \end{pmatrix}
    \begin{pmatrix}
        \Ib_n & \Wb_0 \\ \zerob & \Wb_1^\top
    \end{pmatrix}
=
    \begin{pmatrix}
        \Pb_0 & \Ib_n \\ \Ib_n & \Wb_0
    \end{pmatrix}.
$$
Consequently, the diagonal blocks in~\eqref{eq:W-Problem_dyn} become linear in the variables $\Pb_0$ and $\Wb_0$. The off-diagonal blocks, specified by $\Vbc_\ell^\top \Phib_\ell \Wbc_\ell \Vbc_\ell$, are
$$
\begin{pmatrix}
    \Pb_0 (\Ab\!+\! \Bb \Db \Cb)\!+\!\Pb_1 \Gb \Cb &
    \!\!\Pb_0 \Bb \Eb \!+\! \Pb_1 \Hb\\
     \Ab+\Bb\Db\Cb & 
     \Bb \Eb
\end{pmatrix}\! \begin{pmatrix}
    \Ib & \!\!\Wb_0 \\ \zerob & \!\!\Wb_1^\top
\end{pmatrix}\!.
$$
Now, by introducing the variables $\Lb_0:=\Pb_0\Bb \Db+\Pb_1\Gb$, 
$$
\Lb_1\!:=\!(\Pb_0 (\Ab\!+\!\Bb \Db \Cb)+\!\Pb_1 \Gb \Cb )\Wb_0+(\Pb_0 \Bb \Eb +\! \Pb_1 \Hb) \Wb_1^\top\!,
$$
and $\Lb_2:=\Db \Cb \Wb_0+\Eb \Wb_1^\top$, we obtain
\begin{equation}
    \label{eq:dynamicconvLMI}
\Vbc_\ell^\top \Phib_\ell \Wbc_\ell \Vbc_\ell=\begin{pmatrix}
    \Pb_0 \Ab+ \Lb_0 \Cb &
    \Lb_1 \\
     \Ab+\Bb\Db\Cb & 
     \Ab\Wb_0 + \Bb \Lb_2
\end{pmatrix},
\end{equation}
which is likewise linear in the variables (i.e., $\Pb_0$ and $\Wb_0$ as well as $\Lb_0,\Lb_1,\Lb_2,$ and $\Db$). As a result, 
\eqref{eq:W-Problem_dyn} becomes an LMI. 
Furthermore, the replacements allow for a non-conservative controller parameter reconstruction as long as $\Wb_1$ and $\Pb_1$ have full row rank, which is remarkable. 
Typically, the controller state dimension 
is chosen to be~$n$, which results in a unique reconstruction as follows. Applying Schur's complement to \mbox{$\Vbc_\ell^\top \Wbc_\ell \Vbc_\ell\succ \zerob$} reveals that $\Ib_n-\Wb_0\Pb_0$ is non-singular. Hence, carrying out a singular value decomposition of ${\Ib_n-\Wb_0\Pb_0}$ 
leads to a non-singular $\Wb_1 \Pb_1^\top$.
Based on that, one can first reconstruct $\Gb$ and $\Eb$ from $\Lb_0$ and $\Lb_2$ and subsequently $\Hb$ from $\Lb_1$ (see \cite[p. 88]{scherer_lec} for details).

Next, we discuss whether or not the approach can be adapted for the desired FIR controller design. Clearly, a FIR design is not possible without further ado, since the required structure of $\Hb$ and $\Gb$ as in \eqref{eq:controllerParameters} introduces non-trivial restrictions for the variables $\Lb_0$ and $\Lb_1$. 
A possible adjusting knob to address such issues is a variation of $\Vbc_\ell$. 
Remarkably, the trivial choice $\Vbc_\ell:=\Ib_{n+\ell p}$ 
with the substitutions $\Lb_3:=\Db \Cb \Wb_0+\Cb\Wb_1^\top$ and $\Lb_4:=\Db\Cb\Wb_1^\top+\Cb\Wb_2$ leads to an LMI.
However, similar to the issues coming along with~\eqref{eq:FCK}, 
subsequently finding a compatible $\Db$ 
is usually not possible\footnote{A solution is possible for the previously mentioned special case ${\Cb=\Ib_n}$.\!\!}.
Thus, a key property of~\eqref{eq:dynamicconvLMI} is the linear appearance of $\Db$ (above within $\Ab+\Bb\Db\Cb$) implying that $\Db$ is determined when solving the LMI. 
Based on this observation, suitable transformations parametrized by
$$
    \Vbc_\ell:=
    \begin{pmatrix}
    \Vb_0 & \Vb_1 \\ \Vb_2 & \Vb_3
    \end{pmatrix}
    \qquad \text{and} \qquad 
    \Wbc_\ell \Vbc_\ell =:\begin{pmatrix}
    \Vbt_0 & \Vbt_1 \\ \Vbt_2 & \Vbt_3
    \end{pmatrix},
$$
where the block matrices $\Vb_0,\Vb_1,\Vb_2,$ and $\Vb_3$ 
can either be variables (as $\Pb_i$ in~\eqref{eq:VScherer}) or constants (as $\Ib_n$ and $\zerob$ in ~\eqref{eq:VScherer}), must lead to at least one linear occurrence of $\Db$ in~\eqref{eq:W-Problem_dyn}. 
In the following, let us focus on the off-diagonal terms $\Vbc^\top \Phib_\ell \Wbc_{\ell} \Vbc$, which consist of blocks of the form
\begin{equation*}
     \Vb_i^\top\!(\Ab+\Bb\Db\Cb)\Vbt_j+\Vb_i^\top \!\Bb \Eb \Vbt_k + \Vb_l^\top \Gb \Cb \Vbt_j + \Vb_l^\top\! \Hb \Vbt_k,
\end{equation*}
$(i,j,k,l)\in\{(0,0,2,2),(1,0,2,3),(0,1,3,2),(1,1,3,3)\}$. There,
$\Db$ appears linearly if $\Vb_i$ and $\Vbt_j$ are constant for at least one pair $(i,j)\in\{0,1\}^2$. Consequently, two submatrices in $\Vbc$ are determined. For instance, consider $\Vb_1=\Vbt_0=\Ib_n$ (again assuming the controller state dimension is $n$) as above. Then, two submatrices are fixed via $\Vb_1=\Ib_n$ and 
$\Wb_0\Vb_0+\Wb_1\Vb_2=\Ib_n=\Vbt_0$.
In addition to these restrictions, one has to deal with the nonlinearities within $\Vbc^\top \Phib_\ell \Wbc_{\ell} \Vbc$ as well. Unfortunately, with the previous preparation at hand, a simple argument against the existence of a suitable choice for $\Vbc_\ell$ is then as follows.
Observe that every block in $\Vbc^\top \Phib_\ell \Wbc_{\ell} \Vbc$ needs at least one replacement, since different nonlinear terms occur in each of them. 
However, only three degrees of freedom (including $\Eb$) are left. Thus, neither a consistent replacement nor making the corresponding $\Vb_i, \Vb_l, \Vbt_k,$ and $\Vbt_j$ constant is possible. Thus, the fixed parameters $\Gb$ and $\Hb$ prevent a convexification via~\eqref{eq:W-Problem_dyn}.

\section{Numerical Analysis}
\label{sec:mainnumerics}

The analysis in the previous sections showed that, on the one hand, the existence of a stabilizing FIR controller can easily be checked by testing whether the system is strongly stabilizable (see Section~\ref{sec:maintheory}). 
On the other hand, since classical design approaches based on convex reformulations typically 
fail (see Section~\ref{sec:nofreelunch} and the examples studied below), directly designing a FIR controller appears to be hard.
Yet, there exist various numerical tools tailored for the design of static output feedback which is compatible with a FIR controller via~\eqref{eq:uOutputFeedbackAugmentedY}. This allows us to validate and discuss our results based on selected examples. 
For instance, the HiSyn toolbox~\cite{Popov_HiSyn} provides a useful toolkit. Moreover, we will consider genetic algorithms (GA), which also have been applied successfully to design output feedback \cite{Vesely2005_Genetic_Algorithm_Output_Feedback} and even classical FIR filters \cite{Ababneh_FIR_GA,Suckley_FIR_GA}. Specifically, we are using the GA instance from MATLAB's global optimization toolbox.
Recalling that we aim for stabilizing FIR controllers, we simply consider 
\begin{equation}
    \label{eq:minSpectralRadiusPhi}
 \min_{\Fbc_\ell} \rho\left(\Phib_\ell(\Fbc_{\ell})\right)
\end{equation}
as the goal for the numerical design, where $ \rho\left(\Phib_\ell(\Fbc_{\ell})\right)$ denotes the spectral radius of $\Phib_\ell$, depending on the choice of the controller parameters $\Fbc_\ell$. Note, in this context, that the HiSyn toolbox directly supports the design criterion~\eqref{eq:minSpectralRadiusPhi}. Still, minimizing the spectral radius is non-trivial. 
In fact, it has been pointed out in~\cite{sadabadi2016static} that the non-smooth function~$\rho$ may even have a locally unbounded gradient with respect to $\Fbc_\ell$.
Both characteristics can be handled by the HiSyn toolbox and GA. In fact, the HiSyn toolbox, while using gradient information, treats discontinuities carefully, and GA are gradient-free optimization algorithms by design.

We illustrate our findings and the numerical approaches with four examples. First, we pick up the two examples in \eqref{eq:PIPexamples} (from now on denoted as System~1 and 2, respectively). Then, we consider two examples from the literature.
Specifically, the linearized batch reactor (System~3) in \cite[p.~62]{green2012linear} with a sampling period of $0.1$ is given by
\begin{align}
\nonumber
    &\Ab_{3}=\begin{pmatrix}
    \blind{+}1.18 & 0.00 & \blind{+}0.51 & -0.40 \\
    -0.05 & 0.66 & -0.01 & \blind{+}0.06\\
    \blind{+}0.08 & 0.34 & \blind{+}0.56 & \blind{+}0.38\\
    \blind{+}0.00 & 0.34 & \blind{+}0.09 & \blind{+}0.85\\
    \end{pmatrix}\;\;
    \Bb_{3}=\begin{pmatrix}
    0.00 \\
    0.47 \\
    0.21 \\
    0.21
    \end{pmatrix}
    \\
    \nonumber
    &\Cb_{3}=\begin{pmatrix}
        0 & 1 & 0 & \blind{+}0 \\ 1 & 0 & 1 & -1
    \end{pmatrix}
\end{align}
which serves as a benchmark and is also studied in~\cite{Schlueter2021_ECC}). Lastly, we consider the system investigated for the FIR design in~\cite[p.~5]{Lee2005_FIR} 
(System~4), with 
\begin{align}
\nonumber
&\Ab_{4}=\begin{pmatrix}
1 & -0.3 & 0.6 \\
 0& \blind{+} 0 & 1\\
0.29 & -0.8 & 1\\
\end{pmatrix}\;\;
\Bb_{4}=\begin{pmatrix}
1&0 \\
0&1 \\
1&0 \\
\end{pmatrix}
\\
\nonumber
&\Cb_{4}=\begin{pmatrix}
    1&1&0
\end{pmatrix}.
\end{align}
The selection captures different cases relevant for our problem of interest.
In fact, all systems but System~2 (i.e., $G_2$) are strongly stabilizable. Systems~1 and 4 can be stabilized by classical static output feedback (i.e., a FIR controller with order $\ell=0$), whereas this is impossible for System~2, and it seems to be not possible for System~3.

Now, for all examples and independent of the previous characterization, we repeat the following two steps: i) We investigate whether a convex FIR design is possible based on~\eqref{eq:convexAugmentedOutput}, where we note that considering $\ell=0$ is sufficient according to Lemma~\ref{lem:ConvexInfeasible}; ii)~We apply the HiSyn toolbox as well as GA to (approximately) solve~\eqref{eq:minSpectralRadiusPhi} for increasing $\ell \in \{0,1,\dots,5\}$. 
Regarding step~ii), accounting for the non-convexity of the problem, we perform $10$ optimization runs with randomly selected initial guesses for each $\ell$ and each solver. In this context, if a feasible solution has been found for $\ell-1$, we use the warm start $\Fbc_{\ell}:=\begin{pmatrix} \Fbc_{\ell-1}^\ast & \zerob \end{pmatrix}$ suggested by Lemma~\ref{lem:l+1} as a seed for the random initialization.
\begin{figure}[tb]
 \centering
 \includegraphics{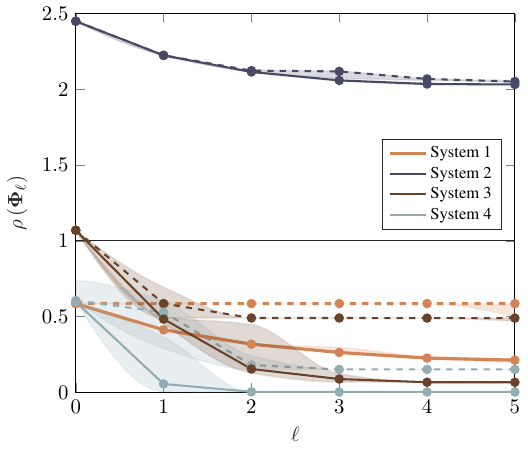}
 \caption{Median closed-loop spectral radius for increasing FIR controller order (HiSyn toolbox in solid, GA in dashed lines). Shaded areas show deviations for different starting values.
 } 
\label{fig:fircontrollers}
\end{figure}
 The central results of our 
 numerical analysis are illustrated in Figure~\ref{fig:fircontrollers} in terms of the optimized spectral radii resulting for the 
 different systems, FIR orders, and solvers. We initially note that the results confirm that a stabilizing FIR controller cannot be found for System~2, as predicted by Theorem~\ref{th:strongstab}. Moreover, it is confirmed that Systems 1 and 4 can be stabilized by classical static output feedback since stabilizing FIR controllers are found even for $\ell=0$. Remarkably, a convex design of such controllers via~\eqref{eq:convexAugmentedOutput} is only possible for System 4 but not for System 1, which highlights the restrictiveness underlying this convexification. 
 At this point, it is interesting to note that the convex approach in \cite{tavazoei2023sufficient}, restricted to SISO systems, is likewise infeasible for System 1 (and 2).
 As expected, the LMI \eqref{eq:convexAugmentedOutput} is further infeasible for Systems~2 and 3 and each $\ell \in \N$. However, as apparent from Figure~\ref{fig:fircontrollers}, a stabilizing FIR controller can be found for System~3 and $\ell\geq 1$ by both the HiSyn toolbox and GA. Finally, one can make two observations.
 First, the spectral radius is typically decreasing with $\ell$ approximately until $\ell=n$ despite the additional design freedom (see Lemma~\ref{lem:l+1}). Second, the HiSyn toolbox provides better or equally good performance, while the running time is about $15$ times faster than GA due to the aforementioned use of gradients in the optimization procedure. 

\section{Conclusion and Outlook}
\label{sec:conclusion}

Motivated by applications in encrypted control, we considered the design of stabilizing FIR controllers in this paper.
We found that the existence of such controllers is closely related to the notion of strong stabilizability (see Section~\ref{sec:maintheory}).
Regarding the design of FIR controllers, one can, in principle, make use of reformulations in terms of augmented static output feedback or structurally restricted dynamic output feedback (Section~\ref{sec:specification}).
However, classical convexifications based on these reformulations suffer from the FIR's special structure, as shown in Section~\ref{sec:nofreelunch}. Therefore, these design approaches are impractical (or even impossible) for a reliable controller synthesis. Nonetheless, designs can, e.g., be achieved by non-convex optimization algorithms as shown in Section~\ref{sec:mainnumerics}. 

Future research directions are twofold. First, we will investigate further classical design approaches in the light of FIR controllers.  
Second, applying FIR controllers in privacy-preserving (encrypted) control as low-level primitives is of interest.

\bibliographystyle{IEEEtran}

\end{document}